# Polarization Dependence of Excess Loss of Amorphous Coating Supermirror in Optical Region for Cavity Ringdown Spectroscopy

Mitsunori Araki[1,†] and Kohsuke Suma,[2,*]

[1]Department of Chemistry, Faculty of Science Division I, Tokyo University of Science, 1-3 Kagurazaka, Shinjuku-ku, Tokyo 162-8601, Japan

[2]Faculty of Education, Kagoshima University 1-20-6 Korimoto, Kagoshima 890-0065, Japan

[†]Current address: Center for Astrochemical Studies, Max-Planck-Institut für extraterrestrische Physik, Giessenbachstrasse 1, Garching bei München, 85748, Germany

∗ Corresponding author

E-mail addresses: araki@mpe.mpg.de (M. Araki), suma@edu.kagoshima-u.ac.jp (K. Suma).

ORCID: 0000-0003-4530-9741(M. Araki), 0009-0000-8232-4532 (K. Suma)

**Abstract**

A long optical path length is critical in achieving sensitive spectroscopy. For cavity ringdown spectroscopy, a cavity consisting of two supermirrors provides a long path length, where high reflectance of the supermirrors results from their slight excess loss. In the case of a crystal coating supermirror, the excess loss has been suggested to depend on polarization. On the other hand, an amorphous coating supermirror was expected to have a negligible polarization dependence. In this work, we measured the excess loss as a function of mirror rotation around its optical axis in the optical region at 681.2 nm for the three amorphous coating supermirrors produced simultaneously by vapor deposition in the same furnace. The back mirror of the cavity was rotated, and the ringdown time as a function of rotational angle was measured every 10 degrees. As a result, sinusoidal variations in excess loss were observed depending on the rotation. The difference in excess loss between the best and worst rotational angles during the rotation of the back mirror reached a maximum of $5.8 \pm 1.2$ ppm. This difference demonstrates the importance of optimizing the rotational angle alignment of supermirrors to achieve high sensitivity via a long path length in cavity ringdown spectroscopy.

## 1. Introduction

An optical cavity is a spatial structure consisting of a pair of highly reflective mirrors for trapping and resonating light, and it has been widely investigated so far, e.g., optical coating [1], scattering [2], and optics mechanics [3]. The optical technology cultivated here has brought a wide range of applications to humanity, exemplified by cutting-edge cases such as the designs of spectroscopic tools and photonic devices [4][5]. For spectroscopy with a cavity, a long optical path length is one of the most critical factors for achieving sensitive detection. In cavity-based techniques such as cavity ringdown (hereafter CRD) spectroscopy, a long path length is obtained by a cavity consisting of two supermirrors, hereafter mirror(s). CRD spectroscopy is a way to measure the pure absorption of gas-phase sample material in the cavity using a laser [6]. An absorbance of the material is evaluated by time-domain behavior, which is a decay curve of pulsed laser light leaking out from the cavity. This method has been widely applied to high-resolution and/or sensitive spectroscopy [7]. The high reflectance of the mirrors is the most critical factor for this spectroscopy. However, despite its importance, the dependence of a mirror's reflectance on a polarization angle of laser light, i.e., dichroism, is not well known.

Recognition of this dependence is meaningful not only in CRD spectroscopy but also in many fields that use an optical cavity. It could be particularly significant in interferometry, such as detecting a gravitational wave, e.g., [8]. The number of round-trip in the interferometers aiming at the gravitational wave achieves three orders of magnitude, e.g., LIGO [9] and Virgo [10]. In this field, mirror coating materials have been investigated by focusing on amorphous optical coatings [9][11].

This dependence arises from birefringence and/or excess loss of the mirrors. Birefringence slightly rotates the polarization angle of laser light, resulting in an apparent change in reflectance when detecting a single polarization component of light. The birefringence-induced dependence was investigated by splitting the two polarization components, vertical and horizontal, with a beam splitter or a polarizer in front of a detector [12][13][14][15]. However, in a setup of usual CRD spectroscopy, both polarization components are simultaneously detected by a detector without beam splitting. This dependence originating from birefringence could not usually affect CRD spectroscopy.

On the other hand, the polarization dependence of excess loss in mirrors is an imperative effect on a CRD spectrometer using a polarized probe laser light, where excess loss is the sum of absorption, transmission, and scattering by the mirrors. This polarization dependence was studied in only two cases in infrared. The polarization dependence at 1.6 μm was reported to be in the order of $10^{-2}$ ppm by Huang & Lehmann in 2008 [16], and that at 4.54 μm was 8.3 ppm by Winkler et al. in 2021 [17]. The latter used the single-crystal GaAs/AlGaAs interference coating mirrors whose crystal axes were fixed parallel, although the former did not specify the type of mirror coating. Winkler et al. explained that the polarization dependence is a behavior of absorption in the mirrors.

The coincidence of the layer boundaries in mirror coating and the nodes of standing waves of light produces the reflection of a supermirror. In the case of the optical region, each coat of the mirror coating is thinner than the infrared region due to the shorter wavelength. Therefore, the dependence of excess loss is presumed to be smaller than that in the infrared region. Additionally, an amorphous coating mirror could have no or minimal polarization dependence of excess loss owing to the random structure of the amorphous coating. Amorphous-coating optical mirrors are assumed to have negligible polarization dependence of excess loss, but the magnitude of the dependence needs to be confirmed. Polarization dependence investigation answers whether rotational-angle adjustment for mirrors is necessary or not to obtain the long path length.

In this study, we report the polarization dependence of excess loss in optical mirrors coated with amorphous material and discuss the necessity of adjusting rotational angles of mirrors to obtain the best sensitivity in cavity ringdown spectroscopy.

## 2. Experimental

We used a standard CRD spectrometer consisting of a tunable pulse laser system and an optical cavity with a length of 83 cm to measure the polarization dependence of excess loss in mirrors, as described in Figure 1. The tunable pulse laser light was taken from a dye laser (LAMBDA Physik, ScanMate 2E) pumped by a Nd:YAG laser (Spectra-Physics, INDI-40-10, 532 nm) running at 10 Hz. The line width of this tunable laser was 0.03 cm$^{-1}$ with an etalon. The dye-laser output was at ~3 mJ/pulse and was inserted into the cavity with the power of ~0.1 mJ/pulse after being attenuated by neutral density filters. This significant attenuation prevented photothermal effects on the surfaces of mirrors. Three aluminum mirrors delivered the laser light to the cavity, which was constructed in the air using a rotatable mirror mount (THORLABS, KS1RS) for the back mirror and a normal mount (Newport, U100-G) for the front mirror. These mounts had canopy structures extending 10 mm from the back mirror and 4 mm from the front mirror, protecting the mirrors from dust. Additionally, a dust cover made of a half-round tube of 6 cm in length was installed in front of the back mirror. A photomultiplier (HAMAMATSU, H10721-20, head-on type, almost no polarization feature) detected a leaked laser light from the cavity, and an oscilloscope (LeCroy, 6050A) observed a ringdown signal.

The transmittance spectrum of a mirror used to make the cavity was observed by a compact spectrometer, as shown in Figure A1. This spectrum suggests that the laser wavelength of 681.2 nm used in this work is in the high reflectance region of the mirrors, as a low transmittance means a high reflectance.

If mirrors are set as windows of a vacuum chamber, atmospheric air pressure causes physical stress in the surface material layer of the mirrors, inducing birefringence. Then, the cavity was constructed in the air to avoid birefringence induced by atmospheric pressure. Furthermore, the back mirror in the rotatable mirror mount was fixed by the minimum and equal pressing force using a retaining ring, giving negligible stress to the mirror.

The signal was transferred to a data acquisition system, including a ringdown calculation function developed with LabVIEW (ver. 2011). Each ringdown event was sampled during 30 μs in 300 data points starting from 2 μs after the head of each ringdown curve. This 2 μs is because the head of the ringdown curve occasionally included the effect of amplified spontaneous emission (ASE), which polluted the exponential decay curve of a laser pulse. Eight ringdown events were averaged, and a ringdown time, τ, was obtained by the least squares fitting to an exponential decay curve. The 40 values of τ were averaged to derive a representative value of τ, and a standard deviation of the 40 values was used as an uncertainty of the representative value. During the measurements, a room air conditioner maintained temperature at 21–22°C and humidity at 21–22%.

A representative raw trace of the ringdown curve is shown in Figure A2 (1). A single-exponential function reproduced this curve well, as Figure A2 (2) describes residuals. If birefringence affected the ringdown curve (see Figures 4b and 4d in [14]), the residuals should contain periodic structure, but this is not the case. Using this setting, 334 ringdown times were continuously recorded, as summarized in the histogram of Figure A3. It was proved that the variation of ringdown times followed a Gaussian distribution as well as Figure 4e of [14]. These two features in Figures A2 and A3 confirmed that single-mode laser operation was achieved, and mode-beating effects, such as polarization, transversal, and birefringence, were mitigated.

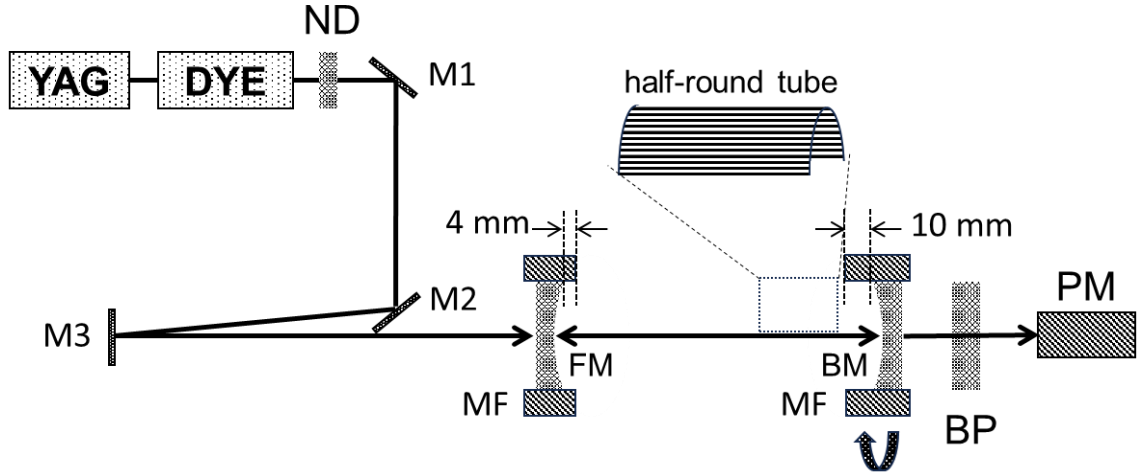

**Figure 1**. Setup of cavity ringdown spectrometer. A spatial filter was not used because it did not improve the ringdown curve in the present wavelength region, i.e., unnecessary modes negligibly appeared in the dye laser. ND: neutral density filters, M1, M2, and M3: aluminum mirrors, FM and BM: front and back supermirrors, MF: mirror folders (MF for BM is rotatable), BP: a bandpass filter of the 680 – 720 nm region, PM: a photomultiplier.

To measure the polarization dependence, we rotated the back mirror of the cavity by a 10-degree step in ascending order around its optical axis. After each rotational step of the back mirror, both the front and back mirrors of the cavity and the aluminum mirror in front of the cavity were re-aligned to obtain the best ringdown time. For reference, if these measurements are tried by randomized mirror angle, i.e., not ascending order, this way needs a long time to re-align the cavity. It doesn't allow measuring many mirror angle points in a limited time. Hence, ascending-order measurement is the ideal way. The diameter of the laser light in the cavity was approximately 3 mm, and the optical axis of the laser light was configured between the physical centers of the front and back mirrors. The contact areas of the laser light and the mirrors did not change significantly during the rotation of the back mirror. Finally, without vignetting, laser light through the cavity was detected in an 8-mm photocathode area of the photomultiplier. A single measurement at a given angle took about 5 minutes, and covering 240 degrees required approximately 2 hours. The measurements for each mirror were continuously executed without the experimenter's rest time because the contact of dust and/or water with mirror surfaces was expected to change the ringdown time. To minimize the effect of dust in the air, we cleaned both the front and back mirrors using methanol and acetone immediately before starting the measurements. In this work, we tested the three mirrors that were simultaneously vapor-deposited in the same furnace, where they were conveniently labeled (a), (b), and (c). The detailed specifications of the mirrors are listed in Table 1. All measurements, except for data in the appendix, were executed in two days to maintain consistency.

### 3. Results and Discussion

Before measuring the polarization dependence of excess loss, the polarization rotation caused by the birefringence of the cavity was investigated by measuring the polarization angles of laser light before and after the cavity using a polarizer. No polarization angle rotation was detected within ±5 degrees, which is consistent with the fact that phase shift per reflection is of the order of microradians or less [16]. Hence, the present setup allows a measurement of the polarization dependence of excess loss with minimal influence from birefringence.

Relations between $\tau$ and the rotational angles of individual mirrors were measured for the three mirrors. Each rotational angle was changed in steps from the smaller to the larger side for more than two hours of the measurements. The values of $\tau$ for all mirrors showed a downward trend along the horizontal axis, as shown in the upper plots of Figure 2. This trend is thought to be an effect of dust depositing from the air or a change of water adsorption on the surface after cleaning with methanol and acetone. Then, this trend would be time-dependent, not angle-dependent. We needed to remove this trend from the observed behavior to clarify the polarization dependence depending on the angle. First, the relations between $\tau$ and the angles were replotted using $1-R$, the average excess loss of the front and back mirrors. The value of $1-R$ is derived from $\tau$ by the relation $\tau = L/\{c(1-R)\}$, where $L$ is the mirror distance (0.83 m), $c$ is the speed of light, and $R$ is the average reflectance of both mirrors. Second, the plots of $1-R$ were least-squares fitted with a first-order line, which described the downtrend as the simplest assumption.

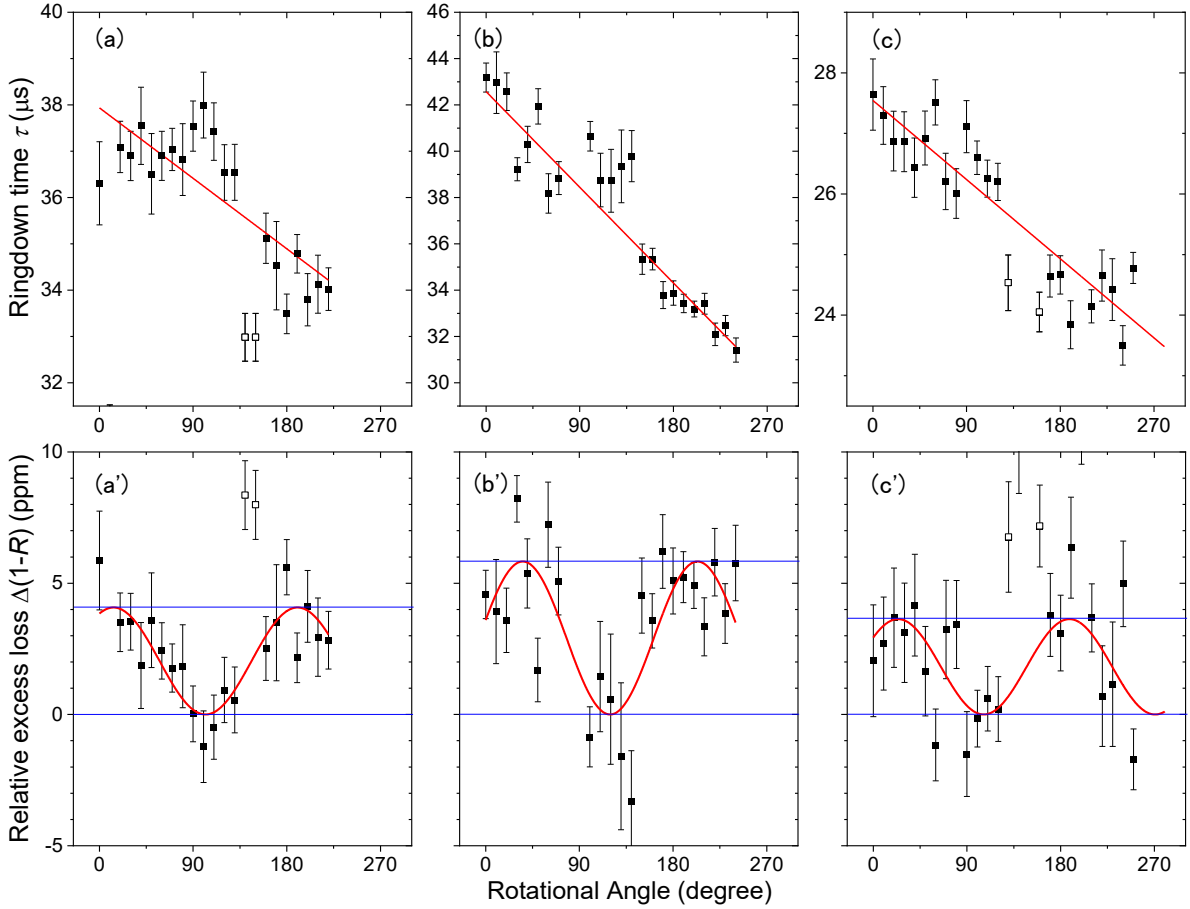

**Figure 2**. Rotational angle dependence of ringdown time and the average excess loss for three amorphous coating mirrors at 681.2 nm.
The upper and lower plots correspond to each other for individual mirrors. Empty squares were not considered in the least squares fitting. The functions used are linear in the upper plots and sinusoidal in the lower plots. The definition of the zero-degree position in the mirror is arbitrary for each. Although all upper plots look divided into two groups, this illusion can be resolved after removing the downtrends and fitting with a sinusoidal curve, as shown in the lower plots. The average excess losses are described in relative values.

After removing the individual downtrends, the plots of $1-R$ for the three mirrors were fitted with a sinusoidal function to find the polarization dependence. The lower plots of Figure 2 show these results using the relative excess loss scale $\Delta(1-R)$. The exceptionally small values of $\tau$ showing more than three ppm larger $\Delta(1-R)$ than the top of the sinusoidal curves were removed from the fitting because it was presumed that they were obtained by insufficient alignment of the mirrors. The three sinusoidal-function cycles in the plots (a'), (b'), and (c') of Figure 2 show the periods of 176.8 ± 13.6, 167.9 ± 12.3, and 164.6 ± 14.1 degrees, respectively, where errors are in 1σ. Therefore, the obtained cycles matched 180 degrees within the 1.1σ errors in all cases of the three mirrors. The repeatability of the sinusoidal-function cycles was confirmed by additional tests with larger rational-angle steps on different days for a mirror (a), as shown in Figure A4. Comparable sinusoidal variations were found in (1) and (2) of this figure despite rapid measurements by the larger steps. These cycles for the three mirrors confirm the detection of polarization dependence. The three sinusoidal-function cycles in Figure 2 show the fitted sinusoidal peak-to-peak values $\Delta(1-R)$, i.e., twice the amplitude, of 4.1 ± 0.6, 5.8 ± 1.2, and 3.6 ± 1.0 ppm, respectively. The standard deviation among the three mirrors was 1.5 ppm, with an average amplitude of 4.5 ppm. This small dispersion would arise from all three mirrors being simultaneously vapor-deposited in the same furnace. This behavior is consistent with the detection of polarization dependence. The relation of $\Delta(1-R)$ to the initial ringdown time $\tau_i$ shows a positive correlation as described in Figure 3, where $\tau_i$ is defined as the $\tau$ value at the zero degree on the first-order line obtained by the fitting. If the measurements of the upper plots in Figure 2 are continued to infinity, $\tau$ asymptotically approaches zero owing to mirror surface contamination by dust and other factors, and $\Delta(1-R)$ also approaches zero, as shown in the extrapolation of the linear fitting of the three data points toward $\tau = 0$ in Figure 3. Hence, this relation of $\Delta(1-R)$ to $\tau_i$ is consistent with the detection of polarization dependence.

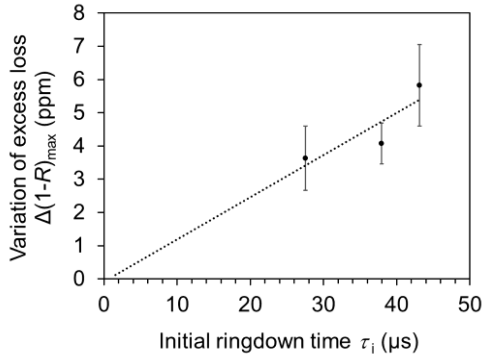

**Figure 3**. Relation of the average-excess-loss variation to the initial ringdown time for the three amorphous coating mirrors. The dotted line shows the linear fitting of the three data points and is extrapolated to the shorter direction of the ringdown time.

Although the mirrors used in this study were coated with amorphous materials, polarization dependence was still observed. This fact might be explained by partial regular patterns caused by crystallization in coating and/or nonuniformity of coating layer thickness. Although the manufacturer did not disclose whether post-deposition annealing was applied, it remains a possible explanation for partial crystallization in the coating materials. Aging is also a possible reason for this crystallization, because these mirrors were made in 2012, and the present data were measured in 2024.

The highest polarization dependence was obtained in the mirror (b) case as $\Delta(1-R) = 5.8 \pm 1.2$ ppm, where the absolute amount of $1-R$ was 64 ppm in $\tau = 64$ μs. This highest dependence corresponds to a 9.1% variation in $\tau$, despite the rotation of only the back mirror. Considering both mirrors of the cavity, the maximum variation of $\tau$ was close to 20%. The path length of a CRD spectrometer is proportional to $\tau$. Rotational-angle optimization for both mirrors is strongly recommended to achieve the long path length, even though amorphous coating mirrors are used in the optical region.

Crystallized structures on the surface of the mirrors were visually investigated in a crossed Nicols configuration with a polarization microscope (MeijiTechno, MT9430). Despite searching at a magnification ratio of 40–100, no crystallized structure was found. This non-detection means that the size of the crystallized area is smaller than 10 μm. Hence, the following model is invalid; crystallized areas oriented in one direction and those oriented at right angles to them exist in roughly equal proportions, with a slight imbalance in their ratio, which accounts for the current small polarization dependence. Since the present polarization dependence of 5.8 ± 1.2 ppm resulted from 5400-time reflections per one side mirror during 30 μs in the CRD system, no detection by one-time transmitting of probe light in this microscope is within the expected. Since this study aims to evaluate the necessity of mirror rotation to achieve the longest ringdown, identifying crystallized structures as a reason for polarization dependence would be a subject for future investigation. An investigation using a further high spatial resolution of a polarization microscope might help find the crystallization of mirrors.

Although the laser wavelength and coating material of mirrors (0.6812 μm, amorphous, back mirror rotation) in this work are different from the case of Winkler et al. [17] (4.54 μm, crystal, both mirror rotation), the polarization dependence is comparable with that of their work (8.3 ppm), as listed in Table 2. On the other hand, the result of Huang & Lehmann [16] (1.6 μm) is smaller than the above two cases. The influence of coating material and a reflection wavelength on the polarization dependence of excess loss would be a future issue.

Aside from the mirrors manufactured by LAYERTEC, we tentatively measured the mirror by Los Gatos Research, which had a similar center wavelength of 690 nm, as a reference. The $\Delta(1-R)$ value of 3.9 ± 1.4 ppm and the period of 189.8 ± 24.0 degrees were found by the 10-point rapid measurements, as shown in Figure A5. This period corresponds with the 180-degree cycle. The sinusoidal curve of this mirror is similar to that of LAYERTEC.

This similarity means that polarization dependence is likely to be common.

In the case of the crystal coating mirror, it was reported that the polarization dependence of excess loss results from absorption [17]. This suggestion is because scattering was negligibly small. Applying this model to the case of the amorphous coating mirrors, the present dependence can be explained by absorption. However, this study aims to judge the necessity of mirror rotation. Although separate evaluation of absorption from transmittance and scattering would also be a future issue, the present results support the necessity.

## 4. Conclusions

This study investigated the polarization dependence of the excess loss in the optical region for the amorphous coating mirrors, aiming to achieve ideal cavity alignment for cavity ringdown spectroscopy. The three mirrors simultaneously produced by vapor deposition in the same furnace were tested at 681.2 nm using the standard CRD spectroscopy setup with the YAG laser-pumped dye laser. The excess loss, depending on the rotational angle of the back mirrors, was measured every 10-degree step for rotation. The excess losses for all three mirrors exhibited sinusoidal variations with a period of approximately 180 degrees. The variation of the excess loss by rotation was $5.8 \pm 1.2$ ppm in the maximum case. These findings suggest that optimizing the rotational angle alignment of mirrors is crucial for achieving higher sensitivity through longer ringdown times, even when using amorphous coating. The insights gained here may contribute to the design of a CRD spectrometer.

Additionally, in the case of a cavity in the gravitational wave interferometry, the number of round-trip is in three orders of magnitude. Since the current measurement results from 5,400 round-trips, the polarization dependence of reflectivity in the interferometry would be smaller than in the present case. However, optimizing the rotational angle alignment of supermirrors to achieve high sensitivity via a long path length would still be crucial. Hence, a CRD measurement might be an excellent benchmark method for evaluating coating materials for interferometers.

**Acknowledgement**
We gratefully acknowledge LAYERTEC GmbH for providing technical information and Prof. Tomoaki Matsui at Kagoshima University for supporting the investigation with the polarization microscope.

**Table 1**

Specification of supermirrors

| | |
|---|---|
| Manufacturer | LAYERTEC [a] |
| Batch number | Z0812002 |
| Center wavelength | 685 nm |
| Coating method | magnetron sputtering |
| Coating materials | $Ta_2O_5/SiO_2$ amorphous |
| Type | plano concave |
| Diameter | 1.0 inch |
| Thickness | 6.35 mm |
| Focal length | 1.1 m |
| Substrate | fused silica |
| Substrate surface accuracy | topside: $\lambda/4$ underside: $\lambda/10$ |
| Manufacturing year | 2012 |

[a] [18]

**Table 2**

Dependence of excess loss on polarization

| | Coating | Wavelength (μm) | Dependence (ppm) |
|---|---|---|---|
| Hung & Lehmann [a] | - | 1.6 | $10^{-2}$ |
| Winkler et al. [b] | crystal | 4.54 | 8.3 [c] |
| This work | amorphous | 0.6812 | 4.5 ± 1.5 (average) |
| | | | 5.8 ± 1.2 (max) |

[a] [16]

[b] [17].

[c] This results from the simultaneous rotation of both mirrors.

**Appendix**

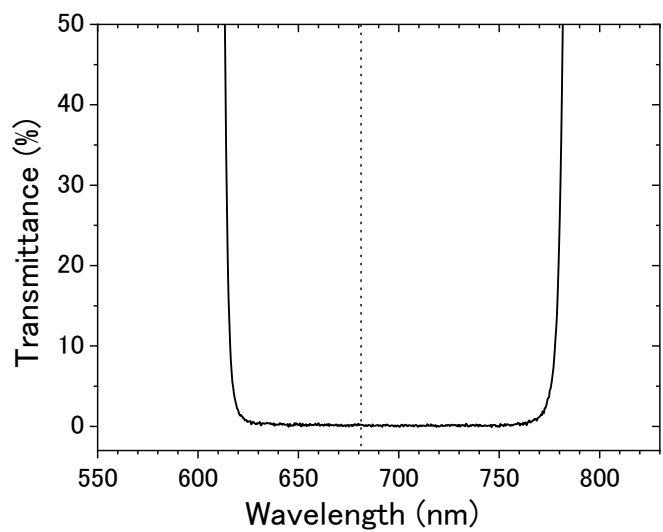

**Figure A1**. Transmittance of supermirror.
The dotted vertical line indicates 681.2 nm of the laser wavelength.

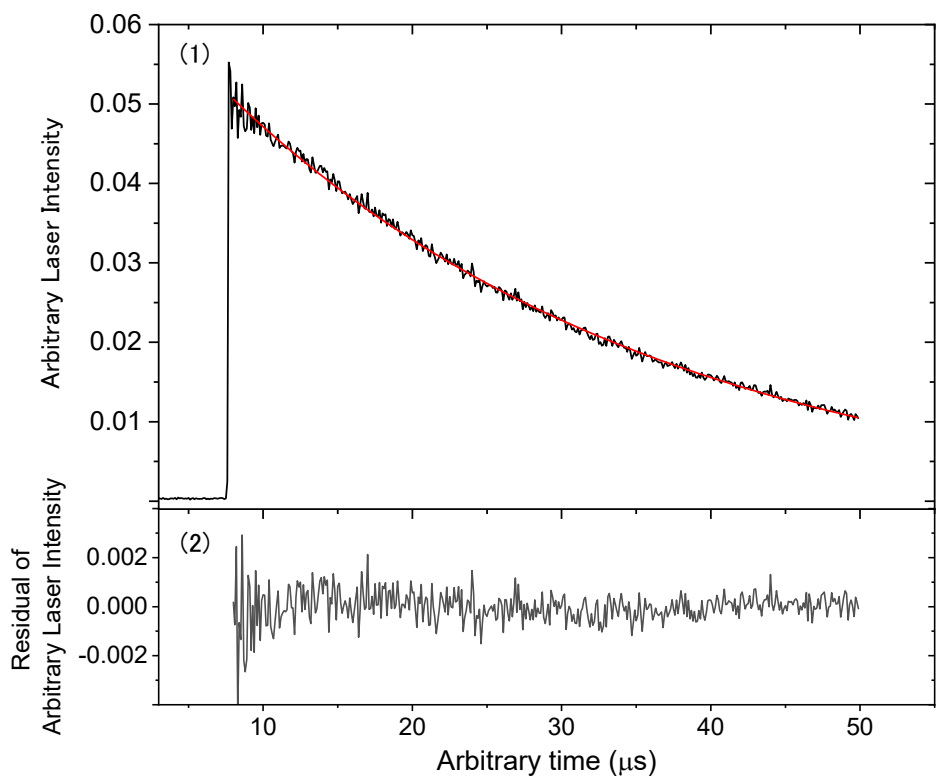

**Figure A2**. Example of a ringdown curve and residuals after single-exponential fitting.
Trace (1) shows the observed ringdown curve in black and its single-exponential fit in red. Trace (2) describes the residuals of the fitting. Aside from the initial few microseconds, the residuals are 1 % of the laser intensity.

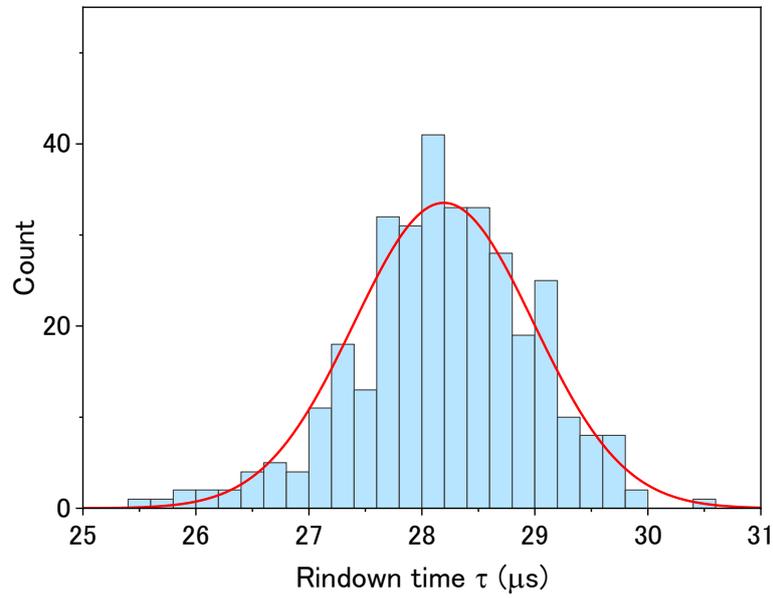

**Figure A3**. Histogram of ringdown times from 334 continuous measurements with a mirror (a) at 681.2 nm.
The average and standard deviation are 28.19 and 0.79 μs, respectively.

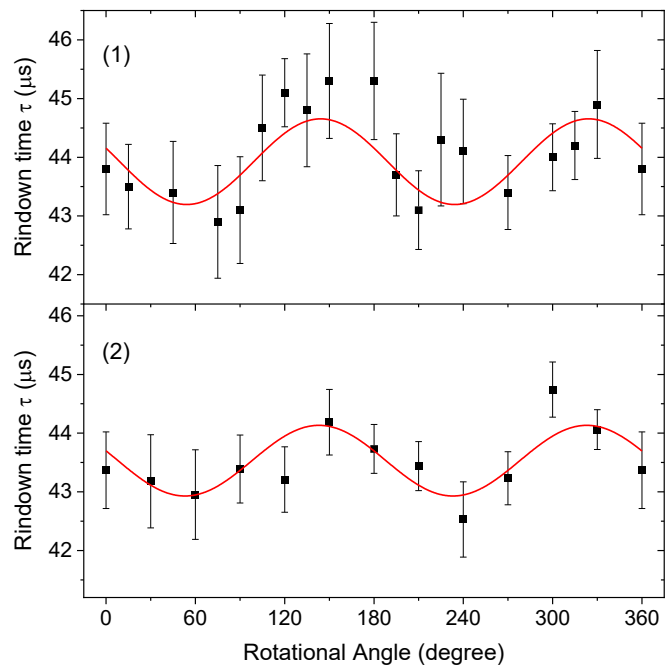

**Figure A4**. Repeatability of the rotational angle dependence of ringdown time for a mirror (a) via rapid measurements with larger rotational angle steps. Both (1) and (2) were measured under accidentally small downtrend conditions. Their measurement days are not the same as each other and as those of Figure 2.

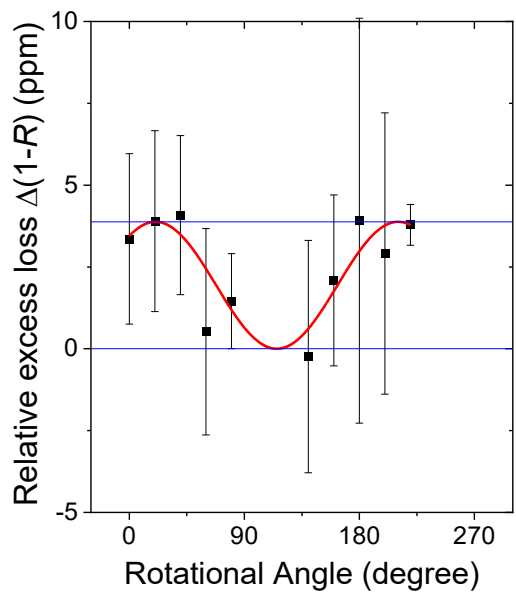

**Figure A5.** Rotational angle dependence of the average excess loss for the Los Gatos Research mirror having a center wavelength of 690 nm via rapid measurements at 681.2 nm.